\documentstyle[twocolumn,prl,aps,amsmath]{revtex}
\tighten

\newcommand{\mint}{- \kern-0.9em \int}

\begin{document}

\draft
\title{Dynamical pattern formation upon dewetting}
\author{J\"urgen Becker and G\"unther Gr\"un}
\address{Institut f\"ur Angewandte Mathematik, Beringstr. 6, D-53115 Bonn, Germany}

\author{Ralf Seemann and Karin Jacobs}
\address{Universit\"at Ulm, Abteilung Angewandte Physik, D-89069 Ulm, Germany}

\date{\today}
\maketitle

\begin{abstract}
Dewetting of thin liquid films is monitored in situ by atomic
force microscopy, results are compared with simulations. The
experimental setting is mimicked as close as possible using the
experimental parameters including the effective interface
potential. Numerics for the thin film equation are based on recently
developed schemes  which are up to now the only methods
convergent in all relevant space dimensions. Temporal evolution
and morphology of experiment and simulation are compared
quantitatively. Our results explain the origin of complex generic
patterns that evolve upon dewetting.
\end{abstract}

\vspace{0.3cm}

For years, experimental \cite{1,2,3,4,5,6,7,Reiter2} and
theoretical \cite{8,9,10,odb,11,12,13,SharmaEvap} groups studied
spinodal dewetting patterns, yet a quantitative comparison of the
results was impossible since the potentials of the `real'
experimental system were unknown. In a recent study \cite{7a},
however, we showed how to get insight into the underlying
potentials by the analysis of dewetting patterns. Hence, it is now
for the first time possible to validate theoretical models also
quantitatively.

Typically, the evolution of thin liquid films is described by
highly nonlinear fourth order degenerate parabolic differential
equations. In recent years, they became the subject both of
analytical and numerical studies in mathematics
\cite{BerFri1990,BP1998,PGB1998,GBW,GruRu1998,GruRu1999c,gru2000,ZhoBe}.
It is one peculiarity of these equations that they admit globally
non-negative solutions. This is proved by means of certain
integral estimates, called entropy estimates in the mathematical
literature. In order to design efficient and convergent numerical
tools, one is therefore interested in schemes that allow for
discrete counterparts of the entropy estimate. This has recently
been accomplished \cite{GruRu1998,GruRu1999c,gru2000} and
therefore the method used in the present paper can be considered
as particularly reliable.

The stability conditions of liquid films on solid substrates can
be described in terms of the effective interface potential $W$
\cite{8,9,Dietrich88,Schick89}, defined as the excess free energy
per unit area needed to bring two surfaces from infinity to a
distance $u$. If there exists a global minimum of $W(u)$ at finite
$u$, the liquid film will not be stable in general; rather, it will dewet
\cite{8,9,10}. Dewetting typically proceeds by the formation of
dry spots, their growth and coalescence, finally leading to a set
of droplets on the substrate \cite{1}. Depending on the sign of
$W''(u)$, different rupture mechanisms are possible. If
$W''(u)<0$, the system spontaneously forms dry spots (`spinodal
dewetting'), where the sites of the holes are correlated
\cite{5,8,7a} reflecting the spinodal wavelength $\lambda_{s}$.
If $W''(u)>0$, a nucleus, e.g. a dust particle, is needed to
induce dewetting.

In our recent study \cite{7a}, we reconstructed the effective
interface potential $W(u)$ by recording the spinodal wavelength as
function of film thickness $u$ and by determining other system
parameters like the contact angle and the liquid equilibrium
layer. The latter two experimental parameters fix the position and
the depth of the global minimum of $W(u)$. With the explicit
expression of $W(u)$, we now perform new numerical simulations of
film rupture. In this Letter, we focus on the comparison of
experimental and numerical results, on the one hand with respect
to the temporal evolution and on the other hand with respect to
special morphologies of patterns upon spinodal dewetting. As we
will demonstrate, the appearance of so-called `satellite holes' is
not a special pattern limited to volatile liquids, as recent work
by Kargupta et al. \cite{SharmaEvap} might suggest, but is a
generic phenomenon of dewetting liquid films.

The liquid used in our experiments was atactic polystyrene (PS)
with a molecular weight of 2 kg/mol (purchased from Polymer Labs,
Church Stratton, UK, M$_w$/M$_m$ = 1.05). This is large enough to
assure non-volatility, but at the same time small enough for the
polymer melt to be approximately a Newtonian fluid. This is
essential for comparison with simulations, in which viscoelastic
effects were neglected. Films were prepared by spin casting a
toluene solution (Selectipur toluene, Merck, Germany) of PS onto
polished, oxidized Si wafers (Silchem GmbH, Freiberg, Germany).
Before coating, the wafers were thoroughly cleaned using standard
procedures \cite{4,7a,Jop2}. The oxide layer and PS film thickness
were determined by ellipsometry (Optrel GdBR, Berlin, Germany).
The surface of the wafers, called `SiO-wafers' below, consists of
amorphous, 191(1)~nm thick silicon oxide \cite{errornote}. On such
a substrate, PS films are unstable up to a film thickness of about
300~nm\cite{Jop2}. The symmetry-breaking mechanism of dewetting
was monitored by atomic force microscopy (AFM) (Nanoscope III,
Digital Instruments, Santa Barbara) in tapping mode$^{TM}$.
Annealing took place on top of the AFM sample holder which enabled
us to follow the dewetting process at a real time scale. We
therefore continuously scanned the sample and recorded one scan in
every 60 s. Scanning parameters were carefully adjusted not to
affect the liquid layer \cite{Jop2}.

Figure 1 depicts on the left side a series of AFM cross sections
of a 3.9(1)~nm thick PS film dewetting a SiO-wafer at an annealing
temperature of $T_a$ = 53~$^{\circ}$C. The film is unstable and
holes are generated due to spinodal dewetting of the film
\cite{7a,Jop2}. The characteristic wavelength $\lambda_{s}$ can
hardly be observed in the cross sections, since the scan size is
too small. Before annealing, the surface of the PS film is smooth,
exhibiting an rms roughness of below 0.3(1)~nm. As revealed from
the online AFM cross sections, first holes are detected after
roughly 1100~s. Looking at the entire (1.5 $\mu$m$)§^2$ AFM scan,
though, holes pop up within a time interval of about 750~s $<$ $t$
$<$ 4000~s, due to slight thickness variations of the polymer
film. Hence, the cross sections shown in Fig.~1a are a kind of
`random sample' for the spinodal break-up of a hole. On the bottom
of the holes, a 1.3~nm thin layer of PS is left, as was measured
by X-ray reflectometry. It is identified with the equilibrium
thickness of PS on SiO \cite{7a}. During the entire dewetting
process, the contact angle of 7.5(5)$^{\circ}$ was found to remain
constant. Due to mass conservation, the material removed from the
inner side of the hole is accumulated at the perimeter of the
hole.

The effective interface potential for that system can be written
as \cite{7a}
\begin{equation}
W(u)= \frac{\epsilon}{u^8} - \frac{A_{SiO}}{12\pi u^2}  \; ,
\label{potential}
\end{equation}
where $\epsilon$ denotes the strength of the short range part of
the potential, $\epsilon = 6.3(1)\cdot10^{-76}$~Jm$^6$, and
$A_{SiO}$ is the Hamaker constant of PS on SiO,
$A_{SiO}=2.2(4)\cdot10^{-20}$~J. The global minimum in $W(u)$ is
at $u$=1.3(1)~nm reflecting the equilibrium thickness of a PS film
on top of SiO-wafer. $W(u)$ enters the equation of motion of the
liquid film,
\begin{equation}
  \label{eq:pde}
  \eta \frac{\partial u}{\partial t} - \mbox{div} (m(u)\nabla p)=0 ,
\end{equation}
via the pressure $p$ which is given by
\begin{equation}
  \label{eq:druck}
  p=-\sigma\Delta u + W'(u) ,
\end{equation}
where $\sigma$ and $\eta$ denote the surface tension and the
viscosity of the film, respectively.
A no-slip condition at the solid-liquid interface entails $m(u)=\frac 13 u^3$.

For numerical purposes it is important to
distinguish between stabilizing and destabilizing terms.
Therefore, we decompose
\begin{equation}
  \label{eq:w}
  W(u)=W_+(u)+W_-(u) ,
\end{equation}
where the stabilizing term $W_+$ and the destabilizing term $W_-$
are given by $\epsilon u^{-8}$ and $-\frac{A_{SiO}}{12 \pi} u^{-2}$, respectively.

In general, smooth solutions of system (\ref{eq:pde})--(\ref{eq:druck}) are not expected to exist.
Instead, we are looking for so called {\it weak solutions} $u,p$ that satisfy
\begin{gather}
  \label{eq:wf1}
  \int_0^T\!\!\! \int_\Omega \eta u_t \phi + \int_0^T\!\!\! \int_\Omega m(u) \nabla p \nabla \phi =0 \\
  \label{eq:wf2}
  \int_0^T\!\!\! \int_\Omega p\psi = \int_0^T\!\!\!\int_\Omega \sigma \nabla u \nabla\psi + \int_0^T\!\!\!\int_\Omega W'(u)\psi
\end{gather}
for all $ \psi,\phi\in C^1((0,T)\times \Omega)$.
Note that smooth solutions are also weak solutions.
Moreover, (\ref{eq:wf1})--(\ref{eq:wf2}) are the starting point to develop finite element schemes (cf. \cite{GruRu1998}).

For simplicity, we sketch the main idea in one spatial dimension:
Let the space interval $\Omega$ be divided in subintervals
$[x_{i-1},x_i]$, $1\le i\le D$ of equal length $h$. Similarly, assume
a time discretization with nodal points $t_k$, $0\le k\le N$  to
be given. We look for  discrete solutions in the linear finite
element space $V^h$. A canonical basis of $V^h$ is given by
functions $\varphi_i$  which are linear on each subinterval and which  satisfy
$\varphi_i(x_j)=\delta_{ij}$. Our discretization is as follows:

{
  \it
  For given $U^0\in V^h$, find a sequence $(U^k,P^k)\in V^h\times V^h, k=0,\dots,N-1$ such that
  \begin{gather}
    \label{eq:disc1}
    \frac{\eta}{\tau_k} (U^{k+1}-U^k,\Phi)_h + (M(U^{k+1})\partial_x P^{k+1},\partial_x \Phi) =0\\
    \label{eq:disc2}
    \begin{split}
      (P^{k+1},\Psi)_h =& \sigma (\partial_x U^{k+1},\partial_x \Psi)\\
      &+ (W'_+(U^{k+1})+W'_-(U^k),\Psi)_h
    \end{split}
  \end{gather}
  for all $\Psi,\Phi \in V^h$.
}

Here, $(.,.)$ denotes the scalar product in the Lebesgue-space $L^2(\Omega)$ and
$(.,.)_h$ is the scalar product of lumped masses, which is an
approximation of $(.,.)$ given by $(\varphi_i,\varphi_i)_h =
\sum_j (\varphi_i,\varphi_j)$ and $(\varphi_i,\varphi_j)_h=0$ if
$i\ne j$. Crucial is the choice of the discrete version $M$ of the
mobility $m$. The right strategy is to define $M$ on each interval
$[x_i,x_{i+1}]$ as
\begin{equation}
  \label{eq:mob}
  M(U)=
  \left\{
    \begin{array}{ll}
      \left(
        \mint_{U(x_i)}^{U(x_{i+1})} \frac{ds}{m(s)}
      \right)^{-1} & \mbox{if } U(x_i)\ne U(x_{i+1})\\
      m(U(x_i)) & \mbox{ otherwise}
    \end{array}
  \right. ,
\end{equation}
where $\mint$ denotes the mean value integral.

The procedure discussed above naturally endows us with
discrete versions of well-known integral estimates which read
\begin{multline}
  \label{eq:energy}
  \eta \int_\Omega \left( \sigma |\nabla u(T)|^2 + W(u(T)) \right)
  + \int_0^T\!\!\! \int_\Omega m(u) |\nabla p|^2\\
  \le \eta \int_\Omega \left( \sigma |\nabla u_0|^2 + W(u_0) \right)
\end{multline}
and
\begin{multline}
  \label{eq:entropy}
  \eta \int_\Omega G(u(T))
  + \sigma \int_0^T\!\!\! \int_\Omega |\Delta u|^2 + W''_+(u) |\nabla u|^2\\
  \le
  \eta \int_\Omega  G(u_0)
  + K \int_0^T \!\!\!\int_\Omega |\nabla u|^2  ,
\end{multline}
where $G$ denotes a second antiderivative of $m^{-1}$.
They are obtained from (\ref{eq:wf1}) by setting $\phi=p$ and
$\phi(x,t)=G'(u(x,t))$, respectively.

Combining these estimates, it can be shown that discrete solutions
are positive and that in addition both $U$ and $P$ converge
strongly with respect to the topology of the
Lebesgue-Sobolev-space $L^2((0,T),H^{1,2}(\Omega))$. Moreover, $U$
converges strongly to $u$ in the H\"older-space $C^{\frac 12,\frac
18}((0,T)\times\Omega)$. \cite{GruRu1998,GruRu1999c}

The right-hand side of Fig. 1 depicts numerical results of thin
film dewetting. The sequence makes use of the effective interface
potential as reconstructed for PS film on top of a SiO-wafer shown
in Eq.~(1), the viscosity $\eta$ = 12000~Pa~s \cite{Herminghaus01}
and the surface tension $\sigma$= 30.8~mN/m. In the numerical
simulation we used 1000 grid points and an initial film of height
3.9 nm with a slightly roughened surface.

The evolution of a hole in the simulation resembles the
experimental AFM scans. The absolute time scale is in the same
order of magnitude and the morphological development of a hole in
experiment and simulation is very similar. Deviations can be
observed especially in the beginning, where the initial conditions
of the film surface play a significant role. In the experiments,
the rupture time varies in a time interval of  750~s $<$ $t$ $<$
4000~s which is comparable to the time scale in the simulation.

If we start the simulation, however, with an `artificial step'
between the equilibrium layer and the original film thickness of
4.9~nm, which is now slightly thicker than before, it
generates a pattern like the one shown on the right-hand side of Fig.~2.
The step transforms rapidly into a heap that relaxes via an
undulation into the original film thickness. With time, the dips
of the undulation grow deeper and the first dip reaches the
substrate after about 400~s. As the other dips follow to break
through to the substrate, a series of consecutive holes appears,
which are generated by a kind of `cascade effect'.

A strikingly similar morphology can be observed experimentally,
too. Fig. 3a-d depict a temporal series of (10~nm)$^2$ AFM scans
of a 4.9~nm thick PS(2k) film on a SiO-substrate. First, single
holes appear that soon grow in size. From a certain size onwards,
they are surrounded by a second row of holes, followed later by a
third one and so on. We term this pattern `satellite holes'. As
compared to the experimental situation reported before, the
film now is 1~nm thicker. The film can still dewet spinodally, yet
the growth time of the amplitude of the preferred mode is about a
factor of three longer \cite{8,12}, therefore dewetting by
nucleation is the quicker way of dewetting. The onset of the
spinodal wavelength yet can be seen by the wavy pattern on the
film matrix, cf. Fig.~3b and c.

A comparison of simulation and experiment is shown in Fig.~2. The
experimental profiles depicted in the left side of Fig.~2 are
horizontal cross sections of the AFM scans of Fig.~3, taken at the
level indicated by the arrow; $t=0$ is defined as the time when
first small holes can be detected in the cross section. Depending
on the direction the cross sections are cutting through the AFM
scan, first satellite holes can be seen in a time interval between
300~s and 1600~s. The time scale of experiment and simulation
hence is in the same order of magnitude. Slight differences are
to be expected, since i) the viscosity of the experimental system
is known only by roughly a factor of two and ii) effects due to
slippage of the polymer melt are not taken into consideration in
the simulation. Yet in both films, an undulatory behavior of the
rim profile can be observed leading to satellite holes. This is
clear evidence that this dewetting pattern does not require a
volatile liquid, but is generic to a spinodally dewetting system.

In a recent study, we showed that an undulatory rim profile can be
observed on polymer films of thicknesses up to at least 60~nm, if
the polymer melt is below the entanglement length
\cite{SeemannPRL2}. There we showed that theoretically a
disturbance in a Newtonian liquid should always decay via an
undulation and that only the viscoelastic properties of the liquid
may stabilize the system. The observation of an undulatory rim
profile in the simulation, where no viscoelastic properties were
considered, is in accordance with this assertion.

A further interesting observation in both experiments and
simulations is that a dip of an undulation only leads to a hole if
the initial film thickness is below about 7~nm. The reasons for
this are speculative at the moment: It is possible that the
driving force for a dip to grow deeper falls below a certain value
for films thicker than 7~nm. It is also possible that there is an
interplay of the dewetting velocity of the three phase contact
line, as set by the global minimum of $W(u)$, and the viscous flow
of the melt \underline{inside} the dewetting front. Further
experiments and new simulations will surely shed more light into
this discussion.

To sum up, we may say that lubrication approximation is an
appropriate method to model the dewetting of liquid films also
beyond the time of the first hole formation. We thus have a
powerful tool to test the influence of single, yet different
parameters onto the dewetting pattern and its temporal evolution.
A successful example was given in the form of the `cascade' effect
leading to satellite holes. It was shown that this is a generic
pattern of a spinodally dewetting film. We
intend to perform numerical studies of the three dimensional system
in a forthcoming paper.

\acknowledgements It is a great pleasure to thank Stephan
Herminghaus for inspiring discussions and critical reading of the
manuscript. We also thank Ralf Blossey and many members of the
priority program `Wetting at Interfaces' for fruitful
collaborations and the German Science Foundation (grant numbers
JA905/1 and GR1693/1) for funding.

\begin{figure}

\caption{Cross sections of a spinodal rupture scenario in a 3.9~nm thick 
PS(2k) film on a SiO-substrate as seen \textit{in situ} by AFM  (left) in
comparison with a simulated one (right).}

\caption{`Cascade effect' leading to satellite holes in
  a 4.9~nm PS(2k) film on a SiO-substrate; AFM cross sections (right) and
  numerical simulation for $\eta$~=~1200~Pa~s (left).}

\caption{Pattern formation in a 4.9~nm PS(2k) film on a
  SiO substrate exhibiting `satellite holes'  a) t~=~0~s, b) 866~s,
  c) 1303~s, and d) 3060~s. Horizontal cross sections of the scans
  are taken at the height of the arrow and are shown at the left side
  of Fig.~2.}

\end{figure}

\end{document}